# The upgrade system of BESIII ETOF with MRPC technology

**Xiaozhuang Wang,**[a,b,c] **Yongjie Sun,**[a,b,1] **Cheng Li,**[a,b,2] **Yuekun Heng,**[b,c] **Zhi Wu,**[b,c] **Ping Cao,**[a,b] **Hongliang Dai,**[b,c] **Xiaolu Ji,**[b,c] **Wenxuan Gong,**[b,c] **Zhen Liu,**[b,c] **Xiaolan Luo,**[b,c] **Weijia Sun,**[a,c] **Siyu Wang,**[a,b] **Yun Wang,**[a,b] **Rongxing Yang,**[a,b] **Mei Ye,**[b,c] **Jinzhou Zhao**[b,c]

[a] *Department of Modern Physics, University of Science and Technology of China(USTC),
   Hefei, Anhui 230026, People's Republic of China*

[b] *State Key Laboratory of Particle Detection and Electronics,
   USTC-IHEP, People's Republic of China*

[c] *Institute of High Energy Physics(IHEP), Chinese Academy of Sciences(CAS),
   Beijing 100049, People's Republic of China*
   *E-mail*: sunday@ustc.edu.cn, licheng@ustc.edu.cn

ABSTRACT: The Beijing Spectrometer III (BESIII) endcap Time-Of-Filght (ETOF) was proposed to upgrade with Multigap Resistive Plate Chamber (MRPC) technology to substitute the current ETOF of scintillator+PMT for extending time resolutin better than 80 ps and enhance the particle identification capability to satisfy the higher precision requirement of physics. The ETOF system including MRPC modules, front end electronics (FEE), CLOCK module, fast control boards and time to digital modules (TDIG), has been designed, constructed and done some experimental tests seperately. Aiming at examining the quality of entire ETOF system and training the operation of all participated parts, a cosmic ray test system was built at the laboratory and underwent about three months to guarantee performance. In this paper the results will be presented indicating that the entire ETOF system works well and satisfies the requirements of the upgrade.

# Contents

## 1. Introduction

The Beijing Spectrometer III(BESIII) [1][2] is a highly precise general-purpose detector designed for high luminosity $e^+e^-$ collisions in the τ-charm energy region. The current BESIII endcap TOF detector's polar angle region is 0.83 <cosθ < 0.96 ( θ as shown in left of Figure 1.) and consists of two disks of 48 pieces of fast plastic scintillator(BC404) and fine-mesh PMT(Hamamatsu R5924)[3]. The time resolution measured is 135 ps for π (1GeV/c) and the momentum range of K/π separation (2σ) is limited up to 1.1 GeV/c[4][5], which cannot satisfy the higher precision requirement of physics analysis nowadays. A possible reason that worsens the time resolution is the scattering particles such as photons hitting on the scintillator with high rate and uncertain positions[6].

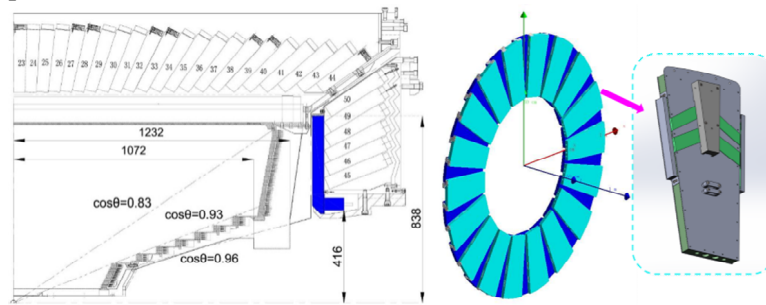

**Figure 1.** (Left)ETOF location at BESIII and (right)MRPC upgrade schematics.

The multi-gap resistive plate chamber (MRPC)[7] was approved at 2012 to substitute BESIII current ETOF detectors with the advantages of excellent time resolution, high detection efficiency, relatively low cost, smaller readout unit and is relatively low sensitivity to neutral particles, which would sufficiently suppress the multi-hit effect at the BESIII endcap region[6]. During the beam tests, the time resolution of π/p @800MeV measured by MRPC could be better 50 ps [8][9] and in predicted simulation [10], the ETOF with MRPC will achieve K/π separation at 95% confidence level up to 1.4 GeV/c.

To examine the quality of the built ETOF system and guarantee the performance before installation a test system under cosmic ray was built at the laboratory. In the following sections, ETOF upgrade system and the cosmic ray test will be presented.

## 2. ETOF upgrade system

### 2.1 Design and structure of MRPC

In the design of final project, each endcap has 36 overlapping MRPC modules as shown in right of Figure 1. ETOF MRPC module has 12 double-end readout strips, with effective inner radius 501 mm and outer radius 822 mm to the beam line. The design of the MRPC module is optimized based on beam test for prototypes in which the single-end and double-end readout schemes are compared[9][10]. In such a design, it has higher granularity than the current scintillator ETOF.

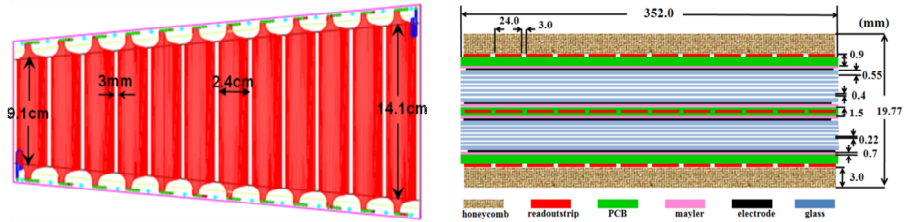

**Figure 2.** (Left)PCB layout of readout strips and (right)cross-section structure of MRPC.

The PCB layout and materials structure of MRPC are shown in the Figure 2. The width of twelve strips are uniform 2.4 cm and lengths are ranging from 9.1 cm to 14.1 cm. The interval between strips is 3 mm. There are total twelve active gas gaps and each gap is 0.22 mm defined by nylon fishing lines. The thicknesses of glass sheets are 0.4 mm and 0.55 mm for the inner and outer sheets respectively and their volume resistivity is around $10^{13}\Omega\cdot$cm. Graphite spray layer is serving as high voltage electrodes on the outer glass sheets surface. The surface resistivity of the graphite spray layer is about 5 MΩ/□. Two pieces of 3 mm thick enforced epoxy honeycomb-boards are attached outmost to ensure the flatness and rigidity of the detector structure. The MRPC module is placed in a gas-tight aluminium box flushed with working gas mixture.

### 2.2 Electronics and gas system

The electronics system of MRPC detectors consists of FEE boards, Calibration-Threshold-Test-Power (CTTP) boards, fast control modules, a CLOCK module and Time-to-Digital (TDIG) convertion modules. They are communicated with and controlled by data acquisition (DAQ) system. The pictures of the electronics modules are shown in Figure 3.

The FEE is designed by four NINO chips developed by the ALICE TOF group [11]. For each channel the timing accuracy is <15 ps RMS when the input charge is > 100 fC. The CTTP board housed in a NIM crate provides power, threshold and test signal for the FEE. The CLOCK module and TDIG module are housed in 9U VME crates. The TDIG modules are based on the HPTDC chips developed by the microelectronics group at CERN. The time resolution of TDIG electronics is approximately 25 ps after integral non-linearity (INL) compensation [12] as shown in left of Figure 4. After receiving and digitizing the signals in TDIG modules, data packing and uploading with predefined format are operated by the data acquisition (DAQ) system via the VME bus. The picture of readout system is shown in right of Figure 4.

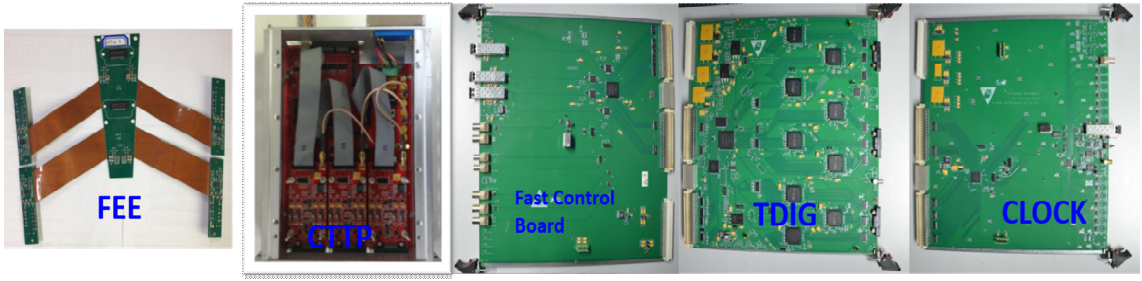

**Figure 3.** The electronics modules of ETOF system: FEE, CTTP, Fast control board, TDIG, CLOCK.

The gas system consists of the gas supply bottles, mass flow controller, flow monitor and gas tanks. The composition of working gas is 90%Freon + 5%$SF_6$ + 5%iso-$C_4H_{10}$. The gas components flow into a buffer tank with component ratio controlled by gas mass flow controller, and then delivered to MRPC modules at a fixed rate of 400 ml/min. To aviod gas moisture copper pipes are selected to connect between gas tanks and MRPC modules while plastic pipes are used gas exhaust lines. Each three MRPC modules are connected serially in a gas circuit.

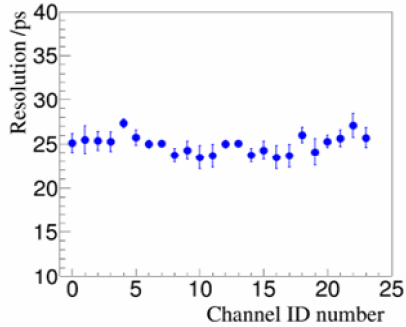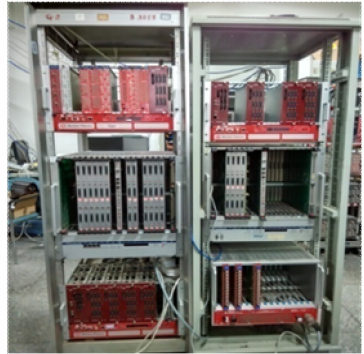

**Figure 4.** (Left)the resolution of electronics and (right) the picture of ETOF system

## 3. Cosmic ray test

### 3.1 The system layout

A cosmic ray test system gathered all parts has been constructed in our laboratory to examine quality of the entire ETOF system. MRPC modules are stacked on four semicircle platforms as shown in left of Figure 5. In order to trigger on the passing particles, 18 pairs of fan shaped scintillator counters, with active area slightly smaller than the active area, are placed above and below the MRPC module respectively. The working voltage is set at $\pm$7000V and FEE threshold is set at 150 mV ( ~ 40fC for input charge[11]). The block diagram of the system setup is illustrated in right of Figure 5.

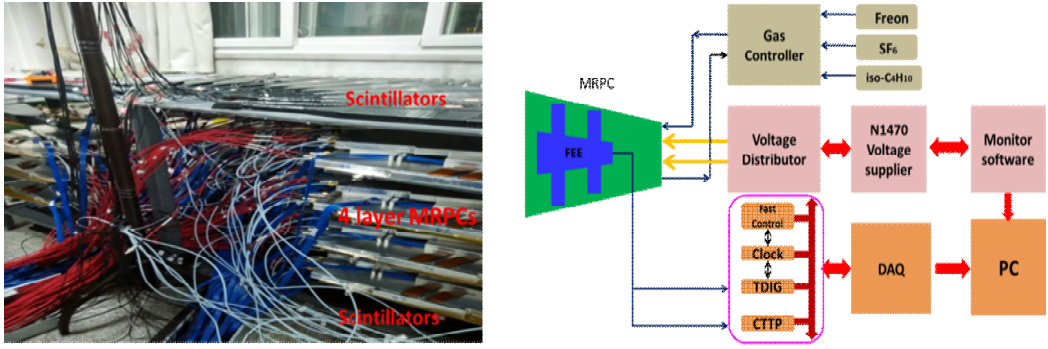

**Figure 5.** (Left)The planting setup of MRPC modules and scintillators and (right) block diagram of test system.

### 3.2 Data analysis

For the convenience of data analysis, the events recorded by the same strip ID are selected for the four vertically stacked MRPC modules. In the analysis, firstly four primary distributions could be achieved by the time difference between one strip and the average of the other three. Considering the T-TOT effect, each time versus TOT distribution with empirical function. After such slewing corrections, four new distributions can be obtained ultimately. The comparisons of TvsTOT and time distributions before and after the corrections can be seen in Figure 6.

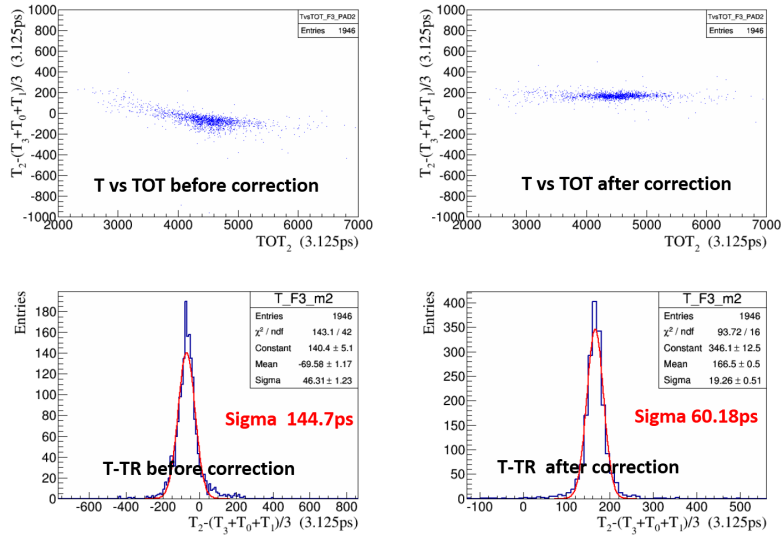

**Figure 6.** The TvsTOT relation and the resolution before and after correction

Define $\sigma_i'$ (i = 1, 2, 3, 4) as measured sigma achieved by fit the corrected time distribution with Gauss function,, while $\sigma_i$ represents the intrinsic time resolution of each MRPC strip. After solving this matrix equation, the intrinsic time resolution of each strip can be obtained by the formula in (3.1):

$$\begin{cases} \sigma_1'^2 = \sigma_1^2 + (\sigma_2^2 + \sigma_3^2 + \sigma_4^2)/9 \\ \sigma_2'^2 = \sigma_2^2 + (\sigma_3^2 + \sigma_4^2 + \sigma_1^2)/9 \\ \sigma_3'^2 = \sigma_3^2 + (\sigma_4^2 + \sigma_1^2 + \sigma_2^2)/9 \\ \sigma_4'^2 = \sigma_4^2 + (\sigma_1^2 + \sigma_2^2 + \sigma_3^2)/9 \end{cases} \Rightarrow \begin{pmatrix} \sigma_1^2 \\ \sigma_2^2 \\ \sigma_3^2 \\ \sigma_4^2 \end{pmatrix} = \begin{pmatrix} 33/32 & -3/32 & -3/32 & -3/32 \\ -3/32 & 32/32 & -3/32 & -3/32 \\ -3/32 & -3/32 & 32/32 & -3/32 \\ -3/32 & -3/32 & -3/32 & 32/32 \end{pmatrix} \begin{pmatrix} \sigma_1'^2 \\ \sigma_2'^2 \\ \sigma_3'^2 \\ \sigma_4'^2 \end{pmatrix}$$

### 3.3 Results

The typical results of resolution and efficiency are shown in Figure 7. The average intrinsic resolution of MRPC strips is approximately 60 ps. The detecting efficiency is determined by the ratio of number of hits in one strip to the number of tracks passing through the strip. The average efficiency of strip is ≥ 96%.

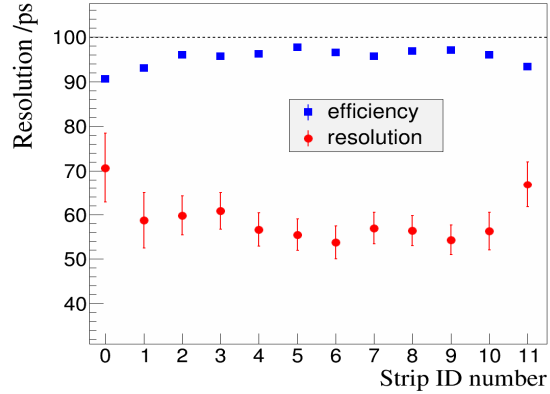

**Figure 7.** Typical resolution and efficiency of MRPC

The noise rates and channel abnormality are also studied in the experiment. In the left of Figure 8, it shows the noise rate per readout channel versus threshold value, which indicates that the noise keeps at low level at normally working threshold rage( >150mV). In the right of Figure 8, it is a typical hits-map example with two abnormal causes during test: high rate and dead channel. High rate situation can be eliminated by shielding the electronics, and the reasons of dead channels probably are no gas flow or HV supply. Both of these two situations can be easily checked out, carefully prevented and effectively solved at laboratory.

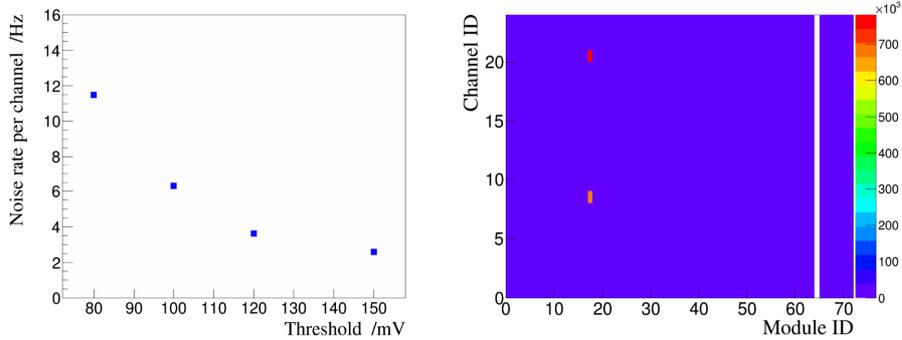

**Figure 8.** (Left)The noise rate per readout channel (right)A typical hits-map example with two abnormal situations

## 4. Conclusions

The upgraded BESIII ETOF system with MRPC technology has been successfully designed, built and examined. The intrinsic time resolution of cosmic ray in MRPC is measured to be approximately 60 ps, significantly better than that of the original plastic scintillator ETOF. The average detecting efficiency for cosmic ray muons is ≥ 96%. The noise rate is quite low and the stability and reliability of the entire system can be guaranteed effectively. These results prove that the performance of the MRPC based ETOF meets the requirements of the upgrade. According to these results, the MRPC ETOF system has been successfully installed at BESIII till October of 2015.

## Acknowledgments

The authors are grateful for the tremendous effort of BESIII ETOF collaboration over the past three years. This work is supported by the National Natural Science Foundation of China (No.11275196).